\documentstyle[12pt,psfig]{article}
\catcode`\"=13\def"#1{\v{#1}}
\def\7{d\llap{\raise0.75ex\hbox{-}}}
\def\8{\rlap{\raise0.5ex\hbox{-}}D}
\def\sub#1{\mbox{\scriptsize\hbox{#1}}}
\newcommand{\sla}[1]{\hbox to 6pt{$#1$\hss{/}}}
\newcommand{\bi}[1]{\hbox{\boldmath{$#1$}}}
\def\half{\textstyle{1\over 2}}
\def\thalf{\textstyle{3\over 2}}
\def\beq{\begin{equation}}
\def\eeq{\end{equation}}
\def\beqa{\begin{eqnarray}}
\def\eeqa{\end{eqnarray}}
\def\b{\!\!\!}
\begin{document}
\baselineskip14pt
\centerline {\large\bf E2 AND C2 AMPLITUDES FOR ELECTROPRODUCTION OF
$\Delta$}

\vspace{28pt}
\centerline {M. FIOLHAIS}

\centerline {\em Department of Physics, University of Coimbra,
3000 Coimbra, Portugal}
\vskip12pt
\centerline{B. GOLLI}

\centerline{\em Faculty of Education and J. Stefan Institute,
University of  Ljubljana, Ljubljana, Slovenia}
\vskip12pt
\centerline{S. \v SIRCA}

\centerline{\em J. Stefan Institute, Jamova 39,
61111 Ljubljana, Slovenia}
\vspace{15pt}

\parbox[h]{6in}{\baselineskip10pt {\small  We compute the
amplitudes for the electromagnetic
$\gamma_{\mbox{\scriptsize v}} N \rightarrow \Delta$ transition
in the framework of soliton models
with quarks and mesons. The ratios E2/M1 and C2/M1 as functions
of the photon four-momentum squared are
dominated by the contribution of the pion cloud and are
in reasonable agreement with the available experimental data.}}

\vspace{30pt}

\noindent The interest in the theoretical calculation of the
electromagnetic amplitudes for the $N
\rightarrow \Delta$ transition  is growing as a consequence of the
recent  and forthcoming
pion electroproduction experiments  in the $\Delta(1232)$
region at the new accelerator
facilities operating with CW electron beams.

At low energies this process can be an important
test of effective models. Here we
consider  two different soliton models
involving quark and meson degrees of
freedom: the linear sigma model (LSM) and the chiral chromodielectric
model (CDM). The nucleon and the $\Delta$ state are
 obtained variationally using the hedgehog ansatz which is
endowed with good quantum numbers of angular momentum and
isospin by means of the Peierls-Yoccoz projection techniques
[1].
The models provide different pictures of the nucleon [2]:
in the LSM there are strong chiral meson fields which are needed
to bind three valence quarks; in the CDM, due to the presence
of the additional chiral singlet $\chi$ field, the three valence
quarks are confined and the strength of the chiral meson
profiles is much smaller.
However, since both
models account for a reasonable description of nucleon properties,
it is interesting to study their predictions
regarding the process $\gamma_{\mbox{\scriptsize v}} N \rightarrow \Delta$,
and try to figure out which
features can be considered  as model independent.
Other calculations of quadrupole amplitudes performed in the framework of
chiral models such as
the CBM [3], the Skyrme model [4] and the NJL
[5] only refer
to photoproduction.

Processes of photo and electroproduction of the $\Delta$
are described
by the vertex

\begin{figure}[h]
\unitlength1.0cm
\begin{center}
\begin{picture}(1.5,2.5)
\put(0.,0.){\psfig{file=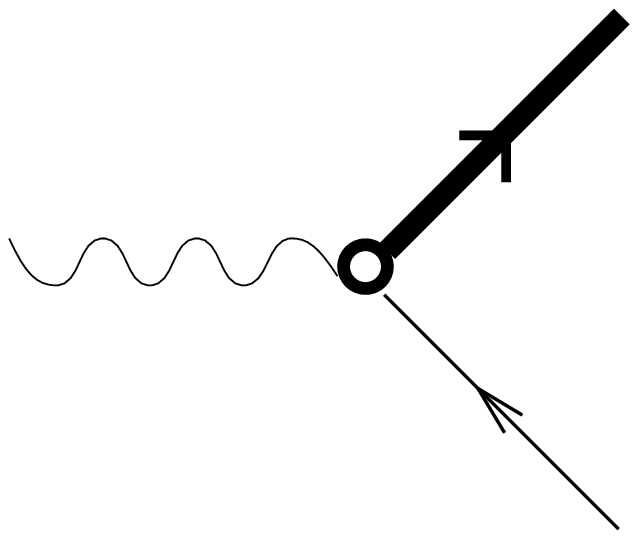,height=3.0cm,angle=-90}}
\put(-0.2,1.9){$\gamma$}
\put(-0.6,0.7){$K=(\omega,{\bi k})$}
\put(3.6,2.3){$\Delta$}
\put(3.6,-0.3){N}
\end{picture}
\end{center}
\label{label}
\end{figure}

\noindent where $\gamma$ is the real (in case of photoproduction) or
the virtual (in case of electroproduction) photon.
In the reference frame in which $\Delta$ is at rest, the
following kinematical expressions hold for the
photon 3-momentum and energy:
\beq
\vert{\bi k}\vert^2=\omega^2-K^2 \, \equiv k^2=
\biggl[ {M_\Delta^2+M_{\sub{N}}^2-K^2\over 2M_\Delta}\biggr]^2-
M_{\sub{N}}^2 \, ; \ \ \ \ \ \
\omega= {M_\Delta^2+K^2-M_{\sub{N}}^2\over2M_\Delta}\>. \label{1}
\eeq

There are only three relevant
multipoles in the expansions  of the $N\rightarrow \Delta$ electromagnetic
transition amplitudes: a transverse electric quadrupole (E2),
a transverse magnetic dipole (M1) and (only in electroproduction)
a  Coulomb quadrupole (C2).  For the models we are
using, both quarks and pions contribute to the electromagnetic
current which is given by
\beq
\widehat{J}^\mu=\overline{q} \gamma^\mu\left(
\frac{1}6+\frac{\tau_3}{2}\right) q
+ (\vec{\pi} \times \partial^\mu \vec{\pi})_3
\equiv (\widehat{\rho}, \widehat{\bi J} ) \, , \label{2}
\eeq
where $\vec \pi$ is the pion field and $q$ is the quark field.

The multipole moments are then given by
\beqa
M^{\sub{M1}}(k) &=&-
{3\over 2}\int \mbox{d}^3{\bi r}\,
\langle\Delta^+  \uparrow\vert
\,(\hat{\bi r}\times\widehat{\bi J})_{\lambda=+1}\,\vert
\mbox{p}  \downarrow
\rangle \,j_1(kr) \label{3} \\
M^{\sub{C2}}(k) &=&-
\sqrt{20\pi}\,\int\mbox{d}^3{\bi r}\,
\langle\Delta^+  \uparrow \vert
\,\widehat{\rho}({\bi r})\,\vert
\mbox{p}   \uparrow \rangle
\,Y_{20}(\hat{\bi r})j_2(kr) \label{4} \\
M^{\sub{E2}'}(k) &=& -\sqrt{10 \pi} {1 \over k}
\int\mbox{d}^3{\bi r}\,\langle\Delta^+  \uparrow \vert
\biggl[ \nabla\times j_2(kr){\bi Y}_{22}^1(\hat{\bi r})\biggr]\cdot
\widehat{\bi J}({\bi r})
 \,\vert\,\mbox{p} \downarrow \rangle\, . \label{5}
\eeqa
Commonly, one uses the current conservation and
writes (\ref{5}) in the form
\beq
  M^{\sub{E2}}(k)
  =  {\sqrt{15\pi}\over 3}\int\mbox{d}^3{\bi r}\,
  \langle\Delta^+ \uparrow \vert\,
  {\omega\over k}\,\widehat{\rho}({\bi r})\,
  {\partial\over\partial r}rj_2(kr)
  - \mbox{i}k\,{\bi r}\cdot\widehat{{\bi J}}({\bi r})\,j_2(kr)
  \,\vert  \mbox{p} \downarrow \rangle\,
  Y_{21}(\hat{{\bi r}})\, .
\label{6}
\eeq
Both expressions (\ref{5}) and (\ref{6})
would lead to the same result if an exact
calculation were performed and $\vert N \rangle$ and
$\vert \Delta \rangle$ were indeed
eigenstates of the model hamiltonians.
In our case the results obtained using (\ref{5})
are very sensitive to the asymptotic
behaviour of the  pion radial profiles.
Since the delta state is described as a soliton and not as a $N\pi$
system at large distances it is better to rely
on (\ref{6}),  similarly to what
has been  done in  other chiral models [3--5].

The matrix elements in eqs. (\ref{3}--\ref{6}) were calculated using  model
states representing the nucleon and the $\Delta$ in the LSM and in the
CDM. The evaluation
is considerably simplified if one takes advantage
of the symmetries of the hedgehog state [6] but  the
final expressions are rather long and therefore are omitted
here.

Assuming the hedgehog ansatz,
the quarks are in s-state and therefore they cannot contribute to
the quadrupole moments (\ref{4}--\ref{6}).
However, through their interactions
with the p-wave
pions they can be excited to d ($j=3/2$) states
in which  case they also contribute.
We have computed the admixture of d-states
in N and $\Delta$ in the CDM and it turns out that
such  excitations  contribute
less than  10\%  to  the total amplitudes [7].
Therefore,  the quadrupole
amplitudes result essentially  from the direct coupling of the photon
to the pion cloud.

In both LSM and CDM we used the standard
choice of parameters for $m_\pi=0.14$ GeV, $m_\sigma=1.2$ GeV
and $f_\pi=0.093\,\mbox{GeV}$. The
simplest version of the CDM  with just a quadratic term in the potential
for the confining field was considered. The relevant parameter
of the model is $G=\sqrt{gM}$ where $M$ is the
$\chi$ mass. Sensible results for the nucleon
properties are obtained for G around 0.2 GeV [8]. For the
LSM the dimensionless constant $g$
should be around 5 [1].
\begin{table}[hbt]
\begin{center}
\begin{tabular}{rccccccccc}
\hline
&&&&&&&\\[-10pt]
Model & $g$ or $G$ & $\mbox{E2/M1}$ & $\mbox{C2/M1}$ &
   $A_{1/2}^{\mbox{\scriptsize q}}$ & $A_{1/2}^\pi$ & $A_{1/2}$ &
   $\mu_{\mbox{\scriptsize iv}}^{\mbox{\scriptsize q}}$&
 $\mu_{\mbox{\scriptsize iv}}^\pi$ &$\mu_{\mbox{\scriptsize iv}}$\\[3pt]\hline
 &&&&&&&\\[-10pt]
&$4.6$&$-1.89$&$-2.35$&$-53$&$-52$&$-105$&$1.22$&$1.16$&$2.38$\\
&&&&&&&\\[-10pt]
LSM&$5.0$&$-1.87$&$-2.33$&$-53$&$-54$&$-107$&$1.20$&$1.20$&$2.40$\\
&&&&&&&\\[-10pt]
&$5.4$&$-1.84$&$-2.29$&$-53$&$-55$&$-108$&$1.17$&$1.22$&$2.39$\\[3pt]\hline
&&&&&&&\\[-10pt]
&$0.18$&$-1.68$&$-2.32$&$-53$&$-18$&$-71$&$1.67$&$0.27$&$1.94$\\
&&&&&&&\\[-10pt]
CDM&$0.19$&$-1.75$&$-2.39$&$-51$&$-20$&$-71$&$1.58$&$0.30$&$1.88$\\
&&&&&&&\\[-10pt]
&$0.20$&$-1.83$&$-2.47$&$-49$&$-21$&$-70$&$1.49$&$0.33$&$1.82$\\[3pt]\hline
&&&&&&&\\[-10pt]
Exp. & & $-1.5\pm 0.4$ & $-$ & & & $-141\pm 5$ &&& $2.35$  \\[3pt]\hline
 \end{tabular}
 \end{center}
 \end{table}

In the Table  we show the E2/M1 and C2/M1 ratios (in \%) at the photon
point for various values of $g$ in LSM and  $G$ (in GeV) in CDM
($g=0.030$ GeV always used).
They are defined by
E2/M1$= M^{\sub{E2}} /  3 \,      M^{\sub{M1}}$ and
C2/M1$= M^{\sub{C2}} /  2\sqrt{2} M^{\sub{M1}}$.
We also present the separate contributions of quarks and pions to
the helicity amplitude $A_{1/2}$ (in units of
$10^{-3}\,\mbox{GeV}^{-1/2}$),
and  the isovector magnetic moment of the nucleon (in n.m.).
The ratio E2/M1 is well reproduced in both models. (The experimental value
quoted was extracted from
[9]; recent measurements in Mainz suggest
a value of around $-2.5\,\%$ [10].)
It is generally assumed that
the photon 3-momentum is sufficiently
small to consider that
${\partial\over\partial r}rj_2(kr)$ in (\ref{6}) can be
replaced by
$3\,j_2(kr)$ and, in such a  case, the ratio E2/M1
reduces to  C2/M1 (we checked that the second term in (\ref{6}),
which is usually neglected [4,5], is indeed very small).
Our results indicate that this reduction
is not justified: in both models, we found that the
contribution to E2/M1, coming from
those terms which are usually dropped,
represents 20--25\% of the total value.
In electroproduction, since $k\ge0.259$ GeV the
ratios E2/M1 and C2/M1 behave quite differently (see bellow).
At the photon point, the experimental value for the helicity amplitude
$A_{1/2}$  is underestimated in both  LSM and in  CDM. The situation is less
favourable in the CDM  due to the
weak pion cloud  pion which leads to a too
small pion contribution  to the amplitude
(the same feature occurs  for the pion part of the nucleon isovector
magnetic moment).
A similar  situation happens regarding the dependence of
$A_{1/2}$ on $K^2$; in the LSM, such  dependence
seems to be more consistent
with  data [7].

The ratios  E2/M1 and C2/M1 obtained in LSM ($g=5.0$)
and CDM ($G=0.2$ GeV) are represented in Fig.~1
as a function of $-K^2$. Experimental points are taken from
[11].

The two ratios are rather insensitive to the details of
the models considered here. As it appears,
the quantity E2/M1 is compatible with
zero for $-K^2 > 0.2$ GeV$^2$, but the ratio C2/M1 is clearly negative,
assuming relatively large values, which may indicate
the presence of an important pion content in the nucleon and the delta.

\vskip2pt
{\sl Acknowledgements}: This work was  supported by the
Calouste Gulbenkian Foundation
(Lisbon), the Ministry of Science of Slovenia, and the European
Commission contract ERB-CIPA-CT-92-2287.
MF would like to acknowledge a travel grant from GTAE (Lisbon).
\newpage

\begin{figure}[t]
\begin{center}
% GNUPLOT: LaTeX picture
\setlength{\unitlength}{0.240900pt}
\ifx\plotpoint\undefined\newsavebox{\plotpoint}\fi
\sbox{\plotpoint}{\rule[-0.200pt]{0.400pt}{0.400pt}}%
\begin{picture}(1125,675)(0,0)
\font\gnuplot=cmr10 at 11pt
\gnuplot
\put(-100,650){\makebox(0,0)[r]{Fig. 1}}
\sbox{\plotpoint}{\rule[-0.200pt]{0.400pt}{0.400pt}}%
\put(220.0,383.0){\rule[-0.200pt]{202.597pt}{0.400pt}}
\put(220.0,113.0){\rule[-0.200pt]{4.818pt}{0.400pt}}
\put(198,113){\makebox(0,0)[r]{--0.15}}
\put(1041.0,113.0){\rule[-0.200pt]{4.818pt}{0.400pt}}
\put(220.0,203.0){\rule[-0.200pt]{4.818pt}{0.400pt}}
\put(198,203){\makebox(0,0)[r]{--0.1}}
\put(1041.0,203.0){\rule[-0.200pt]{4.818pt}{0.400pt}}
\put(220.0,293.0){\rule[-0.200pt]{4.818pt}{0.400pt}}
\put(198,293){\makebox(0,0)[r]{--0.05}}
\put(1041.0,293.0){\rule[-0.200pt]{4.818pt}{0.400pt}}
\put(220.0,383.0){\rule[-0.200pt]{4.818pt}{0.400pt}}
\put(198,383){\makebox(0,0)[r]{0}}
\put(1041.0,383.0){\rule[-0.200pt]{4.818pt}{0.400pt}}
\put(220.0,472.0){\rule[-0.200pt]{4.818pt}{0.400pt}}
\put(198,472){\makebox(0,0)[r]{0.05}}
\put(1041.0,472.0){\rule[-0.200pt]{4.818pt}{0.400pt}}
\put(220.0,562.0){\rule[-0.200pt]{4.818pt}{0.400pt}}
\put(198,562){\makebox(0,0)[r]{0.1}}
\put(1041.0,562.0){\rule[-0.200pt]{4.818pt}{0.400pt}}
\put(220.0,652.0){\rule[-0.200pt]{4.818pt}{0.400pt}}
\put(198,652){\makebox(0,0)[r]{0.15}}
\put(1041.0,652.0){\rule[-0.200pt]{4.818pt}{0.400pt}}
\put(258.0,113.0){\rule[-0.200pt]{0.400pt}{4.818pt}}
\put(258,68){\makebox(0,0){0}}
\put(258.0,632.0){\rule[-0.200pt]{0.400pt}{4.818pt}}
\put(411.0,113.0){\rule[-0.200pt]{0.400pt}{4.818pt}}
\put(411,68){\makebox(0,0){0.2}}
\put(411.0,632.0){\rule[-0.200pt]{0.400pt}{4.818pt}}
\put(564.0,113.0){\rule[-0.200pt]{0.400pt}{4.818pt}}
\put(564,68){\makebox(0,0){0.4}}
\put(564.0,632.0){\rule[-0.200pt]{0.400pt}{4.818pt}}
\put(717.0,113.0){\rule[-0.200pt]{0.400pt}{4.818pt}}
\put(717,68){\makebox(0,0){0.6}}
\put(717.0,632.0){\rule[-0.200pt]{0.400pt}{4.818pt}}
\put(870.0,113.0){\rule[-0.200pt]{0.400pt}{4.818pt}}
\put(870,68){\makebox(0,0){0.8}}
\put(870.0,632.0){\rule[-0.200pt]{0.400pt}{4.818pt}}
\put(1023.0,113.0){\rule[-0.200pt]{0.400pt}{4.818pt}}
\put(1023,68){\makebox(0,0){1}}
\put(1023.0,632.0){\rule[-0.200pt]{0.400pt}{4.818pt}}
\put(220.0,113.0){\rule[-0.200pt]{202.597pt}{0.400pt}}
\put(1061.0,113.0){\rule[-0.200pt]{0.400pt}{129.845pt}}
\put(220.0,652.0){\rule[-0.200pt]{202.597pt}{0.400pt}}
\put(45,382){\makebox(0,0){$\displaystyle{{E2}\over{M1}}$}}
\put(640,13){\makebox(0,0){$-K^2\,[\mbox{GeV}^2]$}}
\put(220.0,113.0){\rule[-0.200pt]{0.400pt}{129.845pt}}
\put(704,336){\circle*{18}}
\put(1002,483){\circle*{18}}
\put(488,350){\circle*{18}}
\put(488,431){\circle*{18}}
\put(602,402){\circle*{18}}
\put(717,417){\circle*{18}}
\put(839,435){\circle*{18}}
\put(1002,458){\circle*{18}}
\put(705,332){\circle*{18}}
\put(602,408){\circle*{18}}
\put(1023,296){\circle*{18}}
\put(294,221){\circle*{18}}
\put(358,490){\circle*{18}}
\put(447,311){\circle*{18}}
\put(567,176){\circle*{18}}
\put(1038,526){\circle*{18}}
\put(704.0,311.0){\rule[-0.200pt]{0.400pt}{12.045pt}}
\put(694.0,311.0){\rule[-0.200pt]{4.818pt}{0.400pt}}
\put(694.0,361.0){\rule[-0.200pt]{4.818pt}{0.400pt}}
\put(1002.0,444.0){\rule[-0.200pt]{0.400pt}{19.031pt}}
\put(992.0,444.0){\rule[-0.200pt]{4.818pt}{0.400pt}}
\put(992.0,523.0){\rule[-0.200pt]{4.818pt}{0.400pt}}
\put(488.0,305.0){\rule[-0.200pt]{0.400pt}{21.681pt}}
\put(478.0,305.0){\rule[-0.200pt]{4.818pt}{0.400pt}}
\put(478.0,395.0){\rule[-0.200pt]{4.818pt}{0.400pt}}
\put(488.0,411.0){\rule[-0.200pt]{0.400pt}{9.636pt}}
\put(478.0,411.0){\rule[-0.200pt]{4.818pt}{0.400pt}}
\put(478.0,451.0){\rule[-0.200pt]{4.818pt}{0.400pt}}
\put(602.0,377.0){\rule[-0.200pt]{0.400pt}{12.045pt}}
\put(592.0,377.0){\rule[-0.200pt]{4.818pt}{0.400pt}}
\put(592.0,427.0){\rule[-0.200pt]{4.818pt}{0.400pt}}
\put(717.0,390.0){\rule[-0.200pt]{0.400pt}{13.009pt}}
\put(707.0,390.0){\rule[-0.200pt]{4.818pt}{0.400pt}}
\put(707.0,444.0){\rule[-0.200pt]{4.818pt}{0.400pt}}
\put(839.0,406.0){\rule[-0.200pt]{0.400pt}{13.731pt}}
\put(829.0,406.0){\rule[-0.200pt]{4.818pt}{0.400pt}}
\put(829.0,463.0){\rule[-0.200pt]{4.818pt}{0.400pt}}
\put(1002.0,379.0){\rule[-0.200pt]{0.400pt}{38.062pt}}
\put(992.0,379.0){\rule[-0.200pt]{4.818pt}{0.400pt}}
\put(992.0,537.0){\rule[-0.200pt]{4.818pt}{0.400pt}}
\put(705.0,305.0){\rule[-0.200pt]{0.400pt}{13.009pt}}
\put(695.0,305.0){\rule[-0.200pt]{4.818pt}{0.400pt}}
\put(695.0,359.0){\rule[-0.200pt]{4.818pt}{0.400pt}}
\put(602.0,384.0){\rule[-0.200pt]{0.400pt}{11.322pt}}
\put(592.0,384.0){\rule[-0.200pt]{4.818pt}{0.400pt}}
\put(592.0,431.0){\rule[-0.200pt]{4.818pt}{0.400pt}}
\put(1023.0,259.0){\rule[-0.200pt]{0.400pt}{18.067pt}}
\put(1013.0,259.0){\rule[-0.200pt]{4.818pt}{0.400pt}}
\put(1013.0,334.0){\rule[-0.200pt]{4.818pt}{0.400pt}}
\put(294.0,172.0){\rule[-0.200pt]{0.400pt}{23.367pt}}
\put(284.0,172.0){\rule[-0.200pt]{4.818pt}{0.400pt}}
\put(284.0,269.0){\rule[-0.200pt]{4.818pt}{0.400pt}}
\put(358.0,436.0){\rule[-0.200pt]{0.400pt}{26.017pt}}
\put(348.0,436.0){\rule[-0.200pt]{4.818pt}{0.400pt}}
\put(348.0,544.0){\rule[-0.200pt]{4.818pt}{0.400pt}}
\put(447.0,275.0){\rule[-0.200pt]{0.400pt}{17.345pt}}
\put(437.0,275.0){\rule[-0.200pt]{4.818pt}{0.400pt}}
\put(437.0,347.0){\rule[-0.200pt]{4.818pt}{0.400pt}}
\put(567.0,113.0){\rule[-0.200pt]{0.400pt}{34.690pt}}
\put(557.0,113.0){\rule[-0.200pt]{4.818pt}{0.400pt}}
\put(557.0,257.0){\rule[-0.200pt]{4.818pt}{0.400pt}}
\put(1038.0,494.0){\rule[-0.200pt]{0.400pt}{15.658pt}}
\put(1028.0,494.0){\rule[-0.200pt]{4.818pt}{0.400pt}}
\put(1028.0,559.0){\rule[-0.200pt]{4.818pt}{0.400pt}}
\put(258,356){\circle*{12}}
\put(258.0,348.0){\rule[-0.200pt]{0.400pt}{3.613pt}}
\put(248.0,348.0){\rule[-0.200pt]{4.818pt}{0.400pt}}
\put(248.0,363.0){\rule[-0.200pt]{4.818pt}{0.400pt}}
\sbox{\plotpoint}{\rule[-0.400pt]{0.800pt}{0.800pt}}%
\put(931,587){\makebox(0,0)[r]{LSM}}
\put(953.0,587.0){\rule[-0.400pt]{15.899pt}{0.800pt}}
\put(258,349){\usebox{\plotpoint}}
\put(258,348.84){\rule{3.854pt}{0.800pt}}
\multiput(258.00,347.34)(8.000,3.000){2}{\rule{1.927pt}{0.800pt}}
\put(274,351.84){\rule{3.614pt}{0.800pt}}
\multiput(274.00,350.34)(7.500,3.000){2}{\rule{1.807pt}{0.800pt}}
\put(289,354.34){\rule{3.614pt}{0.800pt}}
\multiput(289.00,353.34)(7.500,2.000){2}{\rule{1.807pt}{0.800pt}}
\put(304,356.34){\rule{3.614pt}{0.800pt}}
\multiput(304.00,355.34)(7.500,2.000){2}{\rule{1.807pt}{0.800pt}}
\put(319,358.84){\rule{3.854pt}{0.800pt}}
\multiput(319.00,357.34)(8.000,3.000){2}{\rule{1.927pt}{0.800pt}}
\put(335,361.34){\rule{3.614pt}{0.800pt}}
\multiput(335.00,360.34)(7.500,2.000){2}{\rule{1.807pt}{0.800pt}}
\put(350,362.84){\rule{3.614pt}{0.800pt}}
\multiput(350.00,362.34)(7.500,1.000){2}{\rule{1.807pt}{0.800pt}}
\put(365,364.34){\rule{3.854pt}{0.800pt}}
\multiput(365.00,363.34)(8.000,2.000){2}{\rule{1.927pt}{0.800pt}}
\put(381,366.34){\rule{3.614pt}{0.800pt}}
\multiput(381.00,365.34)(7.500,2.000){2}{\rule{1.807pt}{0.800pt}}
\put(396,368.34){\rule{3.614pt}{0.800pt}}
\multiput(396.00,367.34)(7.500,2.000){2}{\rule{1.807pt}{0.800pt}}
\put(411,369.84){\rule{3.614pt}{0.800pt}}
\multiput(411.00,369.34)(7.500,1.000){2}{\rule{1.807pt}{0.800pt}}
\put(426,371.34){\rule{3.854pt}{0.800pt}}
\multiput(426.00,370.34)(8.000,2.000){2}{\rule{1.927pt}{0.800pt}}
\put(442,372.84){\rule{3.614pt}{0.800pt}}
\multiput(442.00,372.34)(7.500,1.000){2}{\rule{1.807pt}{0.800pt}}
\put(457,373.84){\rule{3.614pt}{0.800pt}}
\multiput(457.00,373.34)(7.500,1.000){2}{\rule{1.807pt}{0.800pt}}
\put(472,374.84){\rule{3.854pt}{0.800pt}}
\multiput(472.00,374.34)(8.000,1.000){2}{\rule{1.927pt}{0.800pt}}
\put(488,375.84){\rule{3.614pt}{0.800pt}}
\multiput(488.00,375.34)(7.500,1.000){2}{\rule{1.807pt}{0.800pt}}
\put(503,376.84){\rule{3.614pt}{0.800pt}}
\multiput(503.00,376.34)(7.500,1.000){2}{\rule{1.807pt}{0.800pt}}
\put(518,377.84){\rule{3.614pt}{0.800pt}}
\multiput(518.00,377.34)(7.500,1.000){2}{\rule{1.807pt}{0.800pt}}
\put(533,378.84){\rule{3.854pt}{0.800pt}}
\multiput(533.00,378.34)(8.000,1.000){2}{\rule{1.927pt}{0.800pt}}
\put(549,379.84){\rule{3.614pt}{0.800pt}}
\multiput(549.00,379.34)(7.500,1.000){2}{\rule{1.807pt}{0.800pt}}
\put(564,380.84){\rule{3.614pt}{0.800pt}}
\multiput(564.00,380.34)(7.500,1.000){2}{\rule{1.807pt}{0.800pt}}
\put(579,381.84){\rule{3.854pt}{0.800pt}}
\multiput(579.00,381.34)(8.000,1.000){2}{\rule{1.927pt}{0.800pt}}
\put(610,382.84){\rule{3.614pt}{0.800pt}}
\multiput(610.00,382.34)(7.500,1.000){2}{\rule{1.807pt}{0.800pt}}
\put(625,383.84){\rule{3.854pt}{0.800pt}}
\multiput(625.00,383.34)(8.000,1.000){2}{\rule{1.927pt}{0.800pt}}
\put(595.0,384.0){\rule[-0.400pt]{3.613pt}{0.800pt}}
\put(656,384.84){\rule{3.614pt}{0.800pt}}
\multiput(656.00,384.34)(7.500,1.000){2}{\rule{1.807pt}{0.800pt}}
\put(641.0,386.0){\rule[-0.400pt]{3.613pt}{0.800pt}}
\put(702,385.84){\rule{3.614pt}{0.800pt}}
\multiput(702.00,385.34)(7.500,1.000){2}{\rule{1.807pt}{0.800pt}}
\put(671.0,387.0){\rule[-0.400pt]{7.468pt}{0.800pt}}
\put(763,386.84){\rule{3.614pt}{0.800pt}}
\multiput(763.00,386.34)(7.500,1.000){2}{\rule{1.807pt}{0.800pt}}
\put(717.0,388.0){\rule[-0.400pt]{11.081pt}{0.800pt}}
\put(870,386.84){\rule{3.614pt}{0.800pt}}
\multiput(870.00,387.34)(7.500,-1.000){2}{\rule{1.807pt}{0.800pt}}
\put(778.0,389.0){\rule[-0.400pt]{22.163pt}{0.800pt}}
\put(946,385.84){\rule{3.854pt}{0.800pt}}
\multiput(946.00,386.34)(8.000,-1.000){2}{\rule{1.927pt}{0.800pt}}
\put(885.0,388.0){\rule[-0.400pt]{14.695pt}{0.800pt}}
\put(977,384.84){\rule{3.614pt}{0.800pt}}
\multiput(977.00,385.34)(7.500,-1.000){2}{\rule{1.807pt}{0.800pt}}
\put(962.0,387.0){\rule[-0.400pt]{3.613pt}{0.800pt}}
\put(1007,383.84){\rule{3.854pt}{0.800pt}}
\multiput(1007.00,384.34)(8.000,-1.000){2}{\rule{1.927pt}{0.800pt}}
\put(992.0,386.0){\rule[-0.400pt]{3.613pt}{0.800pt}}
\sbox{\plotpoint}{\rule[-0.500pt]{1.000pt}{1.000pt}}%
\put(931,542){\makebox(0,0)[r]{CDM}}
\multiput(953,542)(20.756,0.000){4}{\usebox{\plotpoint}}
%\put(1019,542){\usebox{\plotpoint}}
\put(258,350){\usebox{\plotpoint}}
\put(258.00,350.00){\usebox{\plotpoint}}
\put(278.33,354.15){\usebox{\plotpoint}}
\put(298.52,358.90){\usebox{\plotpoint}}
\put(318.87,362.97){\usebox{\plotpoint}}
\multiput(319,363)(20.400,3.825){0}{\usebox{\plotpoint}}
\put(339.31,366.57){\usebox{\plotpoint}}
\put(359.78,369.96){\usebox{\plotpoint}}
\put(380.31,372.91){\usebox{\plotpoint}}
\multiput(381,373)(20.573,2.743){0}{\usebox{\plotpoint}}
\put(400.88,375.65){\usebox{\plotpoint}}
\put(421.52,377.70){\usebox{\plotpoint}}
\multiput(426,378)(20.595,2.574){0}{\usebox{\plotpoint}}
\put(442.15,380.01){\usebox{\plotpoint}}
\put(462.82,381.78){\usebox{\plotpoint}}
\put(483.47,383.72){\usebox{\plotpoint}}
\multiput(488,384)(20.710,1.381){0}{\usebox{\plotpoint}}
\put(504.18,385.08){\usebox{\plotpoint}}
\put(524.90,386.00){\usebox{\plotpoint}}
\put(545.63,386.79){\usebox{\plotpoint}}
\multiput(549,387)(20.756,0.000){0}{\usebox{\plotpoint}}
\put(566.38,387.16){\usebox{\plotpoint}}
\put(587.11,388.00){\usebox{\plotpoint}}
\put(607.86,388.00){\usebox{\plotpoint}}
\multiput(610,388)(20.756,0.000){0}{\usebox{\plotpoint}}
\put(628.62,388.00){\usebox{\plotpoint}}
\put(649.35,387.44){\usebox{\plotpoint}}
\put(670.09,387.00){\usebox{\plotpoint}}
\multiput(671,387)(20.710,-1.381){0}{\usebox{\plotpoint}}
\put(690.81,385.70){\usebox{\plotpoint}}
\put(711.52,384.37){\usebox{\plotpoint}}
\multiput(717,384)(20.710,-1.381){0}{\usebox{\plotpoint}}
\put(732.23,382.97){\usebox{\plotpoint}}
\put(752.85,380.68){\usebox{\plotpoint}}
\put(773.49,378.60){\usebox{\plotpoint}}
\multiput(778,378)(20.352,-4.070){0}{\usebox{\plotpoint}}
\put(793.90,374.89){\usebox{\plotpoint}}
\put(814.43,371.91){\usebox{\plotpoint}}
\put(834.78,367.84){\usebox{\plotpoint}}
\multiput(839,367)(20.400,-3.825){0}{\usebox{\plotpoint}}
\put(855.17,363.95){\usebox{\plotpoint}}
\put(875.23,358.61){\usebox{\plotpoint}}
\put(895.09,352.64){\usebox{\plotpoint}}
\put(914.88,346.35){\usebox{\plotpoint}}
\multiput(916,346)(19.271,-7.708){0}{\usebox{\plotpoint}}
\put(934.10,338.55){\usebox{\plotpoint}}
\put(952.99,329.94){\usebox{\plotpoint}}
\put(971.36,320.38){\usebox{\plotpoint}}
\put(988.80,309.14){\usebox{\plotpoint}}
\put(1005.20,296.44){\usebox{\plotpoint}}
\put(1020.89,282.85){\usebox{\plotpoint}}
\put(1023,281){\usebox{\plotpoint}}
\end{picture}
\vspace*{3mm}
% GNUPLOT: LaTeX picture
\setlength{\unitlength}{0.240900pt}
\ifx\plotpoint\undefined\newsavebox{\plotpoint}\fi
\sbox{\plotpoint}{\rule[-0.200pt]{0.400pt}{0.400pt}}%
\begin{picture}(1125,675)(0,0)
\font\gnuplot=cmr10 at 11pt
\gnuplot
\sbox{\plotpoint}{\rule[-0.200pt]{0.400pt}{0.400pt}}%
\put(220.0,383.0){\rule[-0.200pt]{202.597pt}{0.400pt}}
\put(220.0,113.0){\rule[-0.200pt]{4.818pt}{0.400pt}}
\put(198,113){\makebox(0,0)[r]{--0.15}}
\put(1041.0,113.0){\rule[-0.200pt]{4.818pt}{0.400pt}}
\put(220.0,203.0){\rule[-0.200pt]{4.818pt}{0.400pt}}
\put(198,203){\makebox(0,0)[r]{--0.1}}
\put(1041.0,203.0){\rule[-0.200pt]{4.818pt}{0.400pt}}
\put(220.0,293.0){\rule[-0.200pt]{4.818pt}{0.400pt}}
\put(198,293){\makebox(0,0)[r]{--0.05}}
\put(1041.0,293.0){\rule[-0.200pt]{4.818pt}{0.400pt}}
\put(220.0,383.0){\rule[-0.200pt]{4.818pt}{0.400pt}}
\put(198,383){\makebox(0,0)[r]{0}}
\put(1041.0,383.0){\rule[-0.200pt]{4.818pt}{0.400pt}}
\put(220.0,472.0){\rule[-0.200pt]{4.818pt}{0.400pt}}
\put(198,472){\makebox(0,0)[r]{0.05}}
\put(1041.0,472.0){\rule[-0.200pt]{4.818pt}{0.400pt}}
\put(220.0,562.0){\rule[-0.200pt]{4.818pt}{0.400pt}}
\put(198,562){\makebox(0,0)[r]{0.1}}
\put(1041.0,562.0){\rule[-0.200pt]{4.818pt}{0.400pt}}
\put(220.0,652.0){\rule[-0.200pt]{4.818pt}{0.400pt}}
\put(198,652){\makebox(0,0)[r]{0.15}}
\put(1041.0,652.0){\rule[-0.200pt]{4.818pt}{0.400pt}}
\put(258.0,113.0){\rule[-0.200pt]{0.400pt}{4.818pt}}
\put(258,68){\makebox(0,0){0}}
\put(258.0,632.0){\rule[-0.200pt]{0.400pt}{4.818pt}}
\put(411.0,113.0){\rule[-0.200pt]{0.400pt}{4.818pt}}
\put(411,68){\makebox(0,0){0.2}}
\put(411.0,632.0){\rule[-0.200pt]{0.400pt}{4.818pt}}
\put(564.0,113.0){\rule[-0.200pt]{0.400pt}{4.818pt}}
\put(564,68){\makebox(0,0){0.4}}
\put(564.0,632.0){\rule[-0.200pt]{0.400pt}{4.818pt}}
\put(717.0,113.0){\rule[-0.200pt]{0.400pt}{4.818pt}}
\put(717,68){\makebox(0,0){0.6}}
\put(717.0,632.0){\rule[-0.200pt]{0.400pt}{4.818pt}}
\put(870.0,113.0){\rule[-0.200pt]{0.400pt}{4.818pt}}
\put(870,68){\makebox(0,0){0.8}}
\put(870.0,632.0){\rule[-0.200pt]{0.400pt}{4.818pt}}
\put(1023.0,113.0){\rule[-0.200pt]{0.400pt}{4.818pt}}
\put(1023,68){\makebox(0,0){1}}
\put(1023.0,632.0){\rule[-0.200pt]{0.400pt}{4.818pt}}
\put(220.0,113.0){\rule[-0.200pt]{202.597pt}{0.400pt}}
\put(1061.0,113.0){\rule[-0.200pt]{0.400pt}{129.845pt}}
\put(220.0,652.0){\rule[-0.200pt]{202.597pt}{0.400pt}}
\put(45,382){\makebox(0,0){$\displaystyle{{C2}\over{M1}}$}}
\put(640,13){\makebox(0,0){$-K^2\,[\mbox{GeV}^2]$}}
\put(220.0,113.0){\rule[-0.200pt]{0.400pt}{129.845pt}}
\put(704,275){\circle*{18}}
\put(1002,237){\circle*{18}}
\put(488,309){\circle*{18}}
\put(293,244){\circle*{18}}
\put(359,149){\circle*{18}}
\put(446,154){\circle*{18}}
\put(562,323){\circle*{18}}
\put(488,266){\circle*{18}}
\put(602,248){\circle*{18}}
\put(717,266){\circle*{18}}
\put(839,329){\circle*{18}}
\put(1002,235){\circle*{18}}
\put(705,262){\circle*{18}}
\put(602,203){\circle*{18}}
\put(355,154){\circle*{18}}
\put(1030,239){\circle*{18}}
\put(704.0,251.0){\rule[-0.200pt]{0.400pt}{11.322pt}}
\put(694.0,251.0){\rule[-0.200pt]{4.818pt}{0.400pt}}
\put(694.0,298.0){\rule[-0.200pt]{4.818pt}{0.400pt}}
\put(1002.0,206.0){\rule[-0.200pt]{0.400pt}{14.936pt}}
\put(992.0,206.0){\rule[-0.200pt]{4.818pt}{0.400pt}}
\put(992.0,268.0){\rule[-0.200pt]{4.818pt}{0.400pt}}
\put(488.0,271.0){\rule[-0.200pt]{0.400pt}{18.308pt}}
\put(478.0,271.0){\rule[-0.200pt]{4.818pt}{0.400pt}}
\put(478.0,347.0){\rule[-0.200pt]{4.818pt}{0.400pt}}
\put(293.0,188.0){\rule[-0.200pt]{0.400pt}{26.981pt}}
\put(283.0,188.0){\rule[-0.200pt]{4.818pt}{0.400pt}}
\put(283.0,300.0){\rule[-0.200pt]{4.818pt}{0.400pt}}
\put(359.0,126.0){\rule[-0.200pt]{0.400pt}{11.081pt}}
\put(349.0,126.0){\rule[-0.200pt]{4.818pt}{0.400pt}}
\put(349.0,172.0){\rule[-0.200pt]{4.818pt}{0.400pt}}
\put(446.0,131.0){\rule[-0.200pt]{0.400pt}{11.322pt}}
\put(436.0,131.0){\rule[-0.200pt]{4.818pt}{0.400pt}}
\put(436.0,178.0){\rule[-0.200pt]{4.818pt}{0.400pt}}
\put(562.0,300.0){\rule[-0.200pt]{0.400pt}{11.322pt}}
\put(552.0,300.0){\rule[-0.200pt]{4.818pt}{0.400pt}}
\put(552.0,347.0){\rule[-0.200pt]{4.818pt}{0.400pt}}
\put(488.0,228.0){\rule[-0.200pt]{0.400pt}{18.067pt}}
\put(478.0,228.0){\rule[-0.200pt]{4.818pt}{0.400pt}}
\put(478.0,303.0){\rule[-0.200pt]{4.818pt}{0.400pt}}
\put(602.0,221.0){\rule[-0.200pt]{0.400pt}{13.009pt}}
\put(592.0,221.0){\rule[-0.200pt]{4.818pt}{0.400pt}}
\put(592.0,275.0){\rule[-0.200pt]{4.818pt}{0.400pt}}
\put(717.0,224.0){\rule[-0.200pt]{0.400pt}{19.995pt}}
\put(707.0,224.0){\rule[-0.200pt]{4.818pt}{0.400pt}}
\put(707.0,307.0){\rule[-0.200pt]{4.818pt}{0.400pt}}
\put(839.0,293.0){\rule[-0.200pt]{0.400pt}{17.345pt}}
\put(829.0,293.0){\rule[-0.200pt]{4.818pt}{0.400pt}}
\put(829.0,365.0){\rule[-0.200pt]{4.818pt}{0.400pt}}
\put(1002.0,208.0){\rule[-0.200pt]{0.400pt}{13.009pt}}
\put(992.0,208.0){\rule[-0.200pt]{4.818pt}{0.400pt}}
\put(992.0,262.0){\rule[-0.200pt]{4.818pt}{0.400pt}}
\put(705.0,237.0){\rule[-0.200pt]{0.400pt}{12.045pt}}
\put(695.0,237.0){\rule[-0.200pt]{4.818pt}{0.400pt}}
\put(695.0,287.0){\rule[-0.200pt]{4.818pt}{0.400pt}}
\put(602.0,122.0){\rule[-0.200pt]{0.400pt}{39.026pt}}
\put(592.0,122.0){\rule[-0.200pt]{4.818pt}{0.400pt}}
\put(592.0,284.0){\rule[-0.200pt]{4.818pt}{0.400pt}}
\put(355.0,118.0){\rule[-0.200pt]{0.400pt}{17.345pt}}
\put(345.0,118.0){\rule[-0.200pt]{4.818pt}{0.400pt}}
\put(345.0,190.0){\rule[-0.200pt]{4.818pt}{0.400pt}}
\put(1030.0,203.0){\rule[-0.200pt]{0.400pt}{17.345pt}}
\put(1020.0,203.0){\rule[-0.200pt]{4.818pt}{0.400pt}}
\put(1020.0,275.0){\rule[-0.200pt]{4.818pt}{0.400pt}}
\put(258,383){\usebox{\plotpoint}}
\put(258,383){\usebox{\plotpoint}}
\sbox{\plotpoint}{\rule[-0.400pt]{0.800pt}{0.800pt}}%
\put(931,587){\makebox(0,0)[r]{LSM}}
\put(953.0,587.0){\rule[-0.400pt]{15.899pt}{0.800pt}}
\put(258,341){\usebox{\plotpoint}}
\put(258,337.34){\rule{3.400pt}{0.800pt}}
\multiput(258.00,339.34)(8.943,-4.000){2}{\rule{1.700pt}{0.800pt}}
\put(274,333.34){\rule{3.200pt}{0.800pt}}
\multiput(274.00,335.34)(8.358,-4.000){2}{\rule{1.600pt}{0.800pt}}
\put(289,329.34){\rule{3.200pt}{0.800pt}}
\multiput(289.00,331.34)(8.358,-4.000){2}{\rule{1.600pt}{0.800pt}}
\put(304,325.84){\rule{3.614pt}{0.800pt}}
\multiput(304.00,327.34)(7.500,-3.000){2}{\rule{1.807pt}{0.800pt}}
\put(319,322.84){\rule{3.854pt}{0.800pt}}
\multiput(319.00,324.34)(8.000,-3.000){2}{\rule{1.927pt}{0.800pt}}
\put(335,319.84){\rule{3.614pt}{0.800pt}}
\multiput(335.00,321.34)(7.500,-3.000){2}{\rule{1.807pt}{0.800pt}}
\put(350,317.34){\rule{3.614pt}{0.800pt}}
\multiput(350.00,318.34)(7.500,-2.000){2}{\rule{1.807pt}{0.800pt}}
\put(365,314.84){\rule{3.854pt}{0.800pt}}
\multiput(365.00,316.34)(8.000,-3.000){2}{\rule{1.927pt}{0.800pt}}
\put(381,312.34){\rule{3.614pt}{0.800pt}}
\multiput(381.00,313.34)(7.500,-2.000){2}{\rule{1.807pt}{0.800pt}}
\put(396,310.34){\rule{3.614pt}{0.800pt}}
\multiput(396.00,311.34)(7.500,-2.000){2}{\rule{1.807pt}{0.800pt}}
\put(411,307.84){\rule{3.614pt}{0.800pt}}
\multiput(411.00,309.34)(7.500,-3.000){2}{\rule{1.807pt}{0.800pt}}
\put(426,305.34){\rule{3.854pt}{0.800pt}}
\multiput(426.00,306.34)(8.000,-2.000){2}{\rule{1.927pt}{0.800pt}}
\put(442,303.34){\rule{3.614pt}{0.800pt}}
\multiput(442.00,304.34)(7.500,-2.000){2}{\rule{1.807pt}{0.800pt}}
\put(457,301.34){\rule{3.614pt}{0.800pt}}
\multiput(457.00,302.34)(7.500,-2.000){2}{\rule{1.807pt}{0.800pt}}
\put(472,299.34){\rule{3.854pt}{0.800pt}}
\multiput(472.00,300.34)(8.000,-2.000){2}{\rule{1.927pt}{0.800pt}}
\put(488,297.84){\rule{3.614pt}{0.800pt}}
\multiput(488.00,298.34)(7.500,-1.000){2}{\rule{1.807pt}{0.800pt}}
\put(503,296.34){\rule{3.614pt}{0.800pt}}
\multiput(503.00,297.34)(7.500,-2.000){2}{\rule{1.807pt}{0.800pt}}
\put(518,294.34){\rule{3.614pt}{0.800pt}}
\multiput(518.00,295.34)(7.500,-2.000){2}{\rule{1.807pt}{0.800pt}}
\put(533,292.34){\rule{3.854pt}{0.800pt}}
\multiput(533.00,293.34)(8.000,-2.000){2}{\rule{1.927pt}{0.800pt}}
\put(549,290.34){\rule{3.614pt}{0.800pt}}
\multiput(549.00,291.34)(7.500,-2.000){2}{\rule{1.807pt}{0.800pt}}
\put(564,288.84){\rule{3.614pt}{0.800pt}}
\multiput(564.00,289.34)(7.500,-1.000){2}{\rule{1.807pt}{0.800pt}}
\put(579,287.34){\rule{3.854pt}{0.800pt}}
\multiput(579.00,288.34)(8.000,-2.000){2}{\rule{1.927pt}{0.800pt}}
\put(595,285.84){\rule{3.614pt}{0.800pt}}
\multiput(595.00,286.34)(7.500,-1.000){2}{\rule{1.807pt}{0.800pt}}
\put(610,284.34){\rule{3.614pt}{0.800pt}}
\multiput(610.00,285.34)(7.500,-2.000){2}{\rule{1.807pt}{0.800pt}}
\put(625,282.84){\rule{3.854pt}{0.800pt}}
\multiput(625.00,283.34)(8.000,-1.000){2}{\rule{1.927pt}{0.800pt}}
\put(641,281.34){\rule{3.614pt}{0.800pt}}
\multiput(641.00,282.34)(7.500,-2.000){2}{\rule{1.807pt}{0.800pt}}
\put(656,279.34){\rule{3.614pt}{0.800pt}}
\multiput(656.00,280.34)(7.500,-2.000){2}{\rule{1.807pt}{0.800pt}}
\put(671,277.84){\rule{3.614pt}{0.800pt}}
\multiput(671.00,278.34)(7.500,-1.000){2}{\rule{1.807pt}{0.800pt}}
\put(686,276.84){\rule{3.854pt}{0.800pt}}
\multiput(686.00,277.34)(8.000,-1.000){2}{\rule{1.927pt}{0.800pt}}
\put(702,275.34){\rule{3.614pt}{0.800pt}}
\multiput(702.00,276.34)(7.500,-2.000){2}{\rule{1.807pt}{0.800pt}}
\put(717,273.84){\rule{3.614pt}{0.800pt}}
\multiput(717.00,274.34)(7.500,-1.000){2}{\rule{1.807pt}{0.800pt}}
\put(732,272.34){\rule{3.854pt}{0.800pt}}
\multiput(732.00,273.34)(8.000,-2.000){2}{\rule{1.927pt}{0.800pt}}
\put(748,270.84){\rule{3.614pt}{0.800pt}}
\multiput(748.00,271.34)(7.500,-1.000){2}{\rule{1.807pt}{0.800pt}}
\put(763,269.84){\rule{3.614pt}{0.800pt}}
\multiput(763.00,270.34)(7.500,-1.000){2}{\rule{1.807pt}{0.800pt}}
\put(778,268.34){\rule{3.614pt}{0.800pt}}
\multiput(778.00,269.34)(7.500,-2.000){2}{\rule{1.807pt}{0.800pt}}
\put(793,266.84){\rule{3.854pt}{0.800pt}}
\multiput(793.00,267.34)(8.000,-1.000){2}{\rule{1.927pt}{0.800pt}}
\put(809,265.84){\rule{3.614pt}{0.800pt}}
\multiput(809.00,266.34)(7.500,-1.000){2}{\rule{1.807pt}{0.800pt}}
\put(824,264.84){\rule{3.614pt}{0.800pt}}
\multiput(824.00,265.34)(7.500,-1.000){2}{\rule{1.807pt}{0.800pt}}
\put(839,263.34){\rule{3.854pt}{0.800pt}}
\multiput(839.00,264.34)(8.000,-2.000){2}{\rule{1.927pt}{0.800pt}}
\put(855,261.84){\rule{3.614pt}{0.800pt}}
\multiput(855.00,262.34)(7.500,-1.000){2}{\rule{1.807pt}{0.800pt}}
\put(870,260.84){\rule{3.614pt}{0.800pt}}
\multiput(870.00,261.34)(7.500,-1.000){2}{\rule{1.807pt}{0.800pt}}
\put(885,259.84){\rule{3.614pt}{0.800pt}}
\multiput(885.00,260.34)(7.500,-1.000){2}{\rule{1.807pt}{0.800pt}}
\put(900,258.34){\rule{3.854pt}{0.800pt}}
\multiput(900.00,259.34)(8.000,-2.000){2}{\rule{1.927pt}{0.800pt}}
\put(916,256.84){\rule{3.614pt}{0.800pt}}
\multiput(916.00,257.34)(7.500,-1.000){2}{\rule{1.807pt}{0.800pt}}
\put(931,255.84){\rule{3.614pt}{0.800pt}}
\multiput(931.00,256.34)(7.500,-1.000){2}{\rule{1.807pt}{0.800pt}}
\put(946,254.84){\rule{3.854pt}{0.800pt}}
\multiput(946.00,255.34)(8.000,-1.000){2}{\rule{1.927pt}{0.800pt}}
\put(962,253.84){\rule{3.614pt}{0.800pt}}
\multiput(962.00,254.34)(7.500,-1.000){2}{\rule{1.807pt}{0.800pt}}
\put(977,252.34){\rule{3.614pt}{0.800pt}}
\multiput(977.00,253.34)(7.500,-2.000){2}{\rule{1.807pt}{0.800pt}}
\put(992,250.84){\rule{3.614pt}{0.800pt}}
\multiput(992.00,251.34)(7.500,-1.000){2}{\rule{1.807pt}{0.800pt}}
\put(1007,249.84){\rule{3.854pt}{0.800pt}}
\multiput(1007.00,250.34)(8.000,-1.000){2}{\rule{1.927pt}{0.800pt}}
\sbox{\plotpoint}{\rule[-0.500pt]{1.000pt}{1.000pt}}%
\put(931,542){\makebox(0,0)[r]{CDM}}
\multiput(953,542)(20.756,0.000){4}{\usebox{\plotpoint}}
%\put(1019,542){\usebox{\plotpoint}}
\put(258,338){\usebox{\plotpoint}}
\put(258.00,338.00){\usebox{\plotpoint}}
\put(278.12,332.90){\usebox{\plotpoint}}
\put(298.31,328.14){\usebox{\plotpoint}}
\put(318.66,324.07){\usebox{\plotpoint}}
\multiput(319,324)(20.400,-3.825){0}{\usebox{\plotpoint}}
\put(339.05,320.19){\usebox{\plotpoint}}
\put(359.51,316.73){\usebox{\plotpoint}}
\put(380.10,314.11){\usebox{\plotpoint}}
\multiput(381,314)(20.352,-4.070){0}{\usebox{\plotpoint}}
\put(400.51,310.40){\usebox{\plotpoint}}
\put(421.08,307.66){\usebox{\plotpoint}}
\put(441.67,305.04){\usebox{\plotpoint}}
\multiput(442,305)(20.573,-2.743){0}{\usebox{\plotpoint}}
\put(462.24,302.30){\usebox{\plotpoint}}
\put(482.83,299.65){\usebox{\plotpoint}}
\multiput(488,299)(20.573,-2.743){0}{\usebox{\plotpoint}}
\put(503.41,296.95){\usebox{\plotpoint}}
\put(523.98,294.20){\usebox{\plotpoint}}
\put(544.57,291.55){\usebox{\plotpoint}}
\multiput(549,291)(20.573,-2.743){0}{\usebox{\plotpoint}}
\put(565.15,288.85){\usebox{\plotpoint}}
\put(585.73,286.16){\usebox{\plotpoint}}
\put(606.31,283.49){\usebox{\plotpoint}}
\multiput(610,283)(20.573,-2.743){0}{\usebox{\plotpoint}}
\put(626.89,280.76){\usebox{\plotpoint}}
\put(647.47,278.14){\usebox{\plotpoint}}
\put(668.05,275.39){\usebox{\plotpoint}}
\multiput(671,275)(20.352,-4.070){0}{\usebox{\plotpoint}}
\put(688.46,271.69){\usebox{\plotpoint}}
\put(708.97,268.61){\usebox{\plotpoint}}
\put(729.46,265.34){\usebox{\plotpoint}}
\multiput(732,265)(20.400,-3.825){0}{\usebox{\plotpoint}}
\put(749.88,261.62){\usebox{\plotpoint}}
\put(770.23,257.55){\usebox{\plotpoint}}
\put(790.40,252.69){\usebox{\plotpoint}}
\multiput(793,252)(20.400,-3.825){0}{\usebox{\plotpoint}}
\put(810.72,248.54){\usebox{\plotpoint}}
\put(830.78,243.19){\usebox{\plotpoint}}
\put(850.69,237.35){\usebox{\plotpoint}}
\multiput(855,236)(19.690,-6.563){0}{\usebox{\plotpoint}}
\put(870.40,230.87){\usebox{\plotpoint}}
\put(889.99,224.01){\usebox{\plotpoint}}
\put(909.34,216.50){\usebox{\plotpoint}}
\put(928.03,207.58){\usebox{\plotpoint}}
\multiput(931,206)(18.314,-9.767){0}{\usebox{\plotpoint}}
\put(946.34,197.81){\usebox{\plotpoint}}
\put(964.25,187.35){\usebox{\plotpoint}}
\put(980.86,174.91){\usebox{\plotpoint}}
\put(996.59,161.41){\usebox{\plotpoint}}
\put(1011.14,146.61){\usebox{\plotpoint}}
\put(1023,134){\usebox{\plotpoint}}
\end{picture}
\end{center}
\end{figure}

\vspace*{-10mm}

\noindent{\bf References}
\begin{itemize}
\item[1.]
  B. Golli and M. Rosina, {\em Phys. Lett.}  {\bf 165B}, 347 (1985).
  M. C. Birse, {\em Phys. Rev.} {\bf 33D}, 1934 (1986).
\item[2.]
  M. C. Birse, {\em Progr. Part. Nucl. Phys.} {\bf 25}, 1(1990).
\item[3.]
 G. K\"albermann and J. M. Eisenberg, {\em Phys. Rev.} {\bf 28D}, 71 (1983).
 K. Bermuth {\em et al.}, {\em Phys. Rev.}  {\bf 37D}, 89 (1988).
\item[4.]
  A. Wirzba and W. Weise, {\em Phys. Lett.}  {\bf 188B}, 6 (1987).
\item[5.]
 T. Watabe, Chr. V. Christov and K. Goeke, {\em Phys. Lett.}
            {\bf 349B}, 197 (1995).
\item[6.]
  M. \v{C}ibej, M. Fiolhais, B. Golli and M. Rosina,
  {\em J. Phys.}  {\bf 18G}, 49 (1992).
\item[7.]
  M. Fiolhais, B. Golli and S. \v{S}irca,
  preprint submitted to {\em Phys. Lett.~B}.
\item[8.]
  A. Drago, M. Fiolhais and V. Tambini,
  {\em Nucl. Phys.}  {\bf 588A}, 801 (1995).
\item[9.]
  Particle Data Group,
  {\em Phys. Rev.}   {\bf 50D}, 1712 (1994).
\item[10.]
  R. Beck, these Proceedings.
\item[11.]
  Particle Data Group,
  {\em Rev. Mod. Phys.} {\bf 48}, s157 (1976).
\end{itemize}

\end{document}